\documentclass[manuscript,screen,review=false,format=manuscript]{acmart}
\AtBeginDocument{%
  }

\usepackage{amsmath}
\usepackage{amsfonts}
\usepackage{graphicx}
\usepackage{textcomp}
\usepackage{url}
\usepackage{booktabs}
\usepackage{stfloats}
\usepackage{float}
\usepackage[normalem]{ulem}
\usepackage{changepage}
\usepackage[normalsize]{subfigure}
\usepackage{multirow}
\usepackage{paralist}
\usepackage{balance}
\usepackage{breqn}
\usepackage{makecell}
\usepackage{enumitem}
\usepackage{multicol}
\usepackage{bm}
\usepackage[ruled,linesnumbered]{algorithm2e}
\usepackage{colortbl}
\usepackage[normalem]{ulem}

\usepackage{tikz}
\usetikzlibrary{positioning, fit, arrows.meta, backgrounds, calc}

\acmJournal{TOSEM}
\begin{document}

\title{Model-Driven Discipline for Multi-Agent LLMs: Requirement-to-Verification Generation of Traceable System Models}

\author{Ran Wei}
\orcid{0000-0003-2191-1359}
\affiliation{%
  \institution{Lancaster University}
  \city{Lancaster}
  \country{UK}
}
\email{r.wei5@lancaster.ac.uk}

\author{Le Zhu}
\authornote{Corresponding Author}
\email{l-zhu19@mails.tsinghua.edu.cn}
\orcid{0009-0009-8762-0619}
\affiliation{%
  \institution{Tsinghua University}
  \city{Beijing}
  \country{China}
}

\author{Haochi Wang}
\orcid{0009-0005-3768-5335}
\email{wanghaochi1222@gmail.com}
\affiliation{%
  \institution{Harbin Institute of Technology}
  \city{Harbin}
  \country{China}}

\author{Ruizhe Yang}
\orcid{0009-0007-8068-6936}
\email{ruizheyang@mail.dlut.edu.cn}
\affiliation{%
  \institution{Dalian University of Technology}
  \city{Dalian}
  \country{China}}

\author{Jiapeng Guan}
\orcid{0009-0006-6850-8785}
\email{jiapengguan9999@gmail.com}
\affiliation{%
  \institution{Southeast University}
  \city{Nanjing}
  \country{China}
}

\author{Siyuan Ji}
\orcid{0000-0001-6139-3539}
\email{s.ji@lboro.ac.uk}
\affiliation{%
  \institution{Loughborough University}
  \city{Loughborough}
  \country{UK}
}


\author{Yuchen Hu}
\orcid{0009-0001-4169-602X}
\email{yuchen_hu@seu.edu.cn}
\author{Zhe Jiang}
\email{zhejiang.uk@gmail.com}
\orcid{0000-0002-8509-3167}
\affiliation{%
  \institution{Southeast University}
  \city{Nanjing}
  \country{China}
}

\author{Xiangyang Ji}
\email{xyji@tsinghua.edu.cn}
\orcid{0000-0001-9542-5260}
\affiliation{%
  \institution{Tsinghua University}
  \city{Beijing}
  \country{China}
}

\renewcommand{\shortauthors}{R. Wei, L. Zhu, H. Wang, et al.}

\begin{abstract}

Software complexity is a long-standing challenge for system engineers.
Model-Driven Engineering (MDE) addresses it by treating models as first-class artefacts, but a typical MDE process spans many tools and produces heterogeneous models of different system aspects, making traceability, maintenance, and change management difficult.

We propose RADIANT, an engineering methodology that combines MDE with Multi-Agent Large Language Models (LLMs) for complete model-based system development, with a focus on safety-critical systems.
From a carefully specified requirement model, RADIANT automatically generates heterogeneous models across engineering phases---a concept model, a domain-specific modelling language, a conforming system model, and a behaviour model---together with executable, element-level traceability links, on top of which it provides exact, automated change-impact analysis.
Generated behaviour models are translated into CSP and formally verified (e.g.\ for deadlock freedom and convergence) with a counterexample-driven repair loop.
Evaluating RADIANT across three LLMs, we find that the multi-agent decomposition reliably improves the \emph{syntactic validity} of the generated formal artefacts over a single-agent baseline---and their \emph{executability} where the model's code generation permits---while gains in semantic accuracy are model-dependent.
A six-participant study shows an order-of-magnitude ($10$--$15\times$) reduction in development time, and the unmodified pipeline transfers to a second domain.

\end{abstract}

\ccsdesc[500]{Software and its engineering~Software design engineering}

\keywords{Model Driven Engineering, Software Development Methodology, Multi-Agent Large Language Model Framework}

\maketitle

\section{Introduction}
\label{sec:introduction}
Software complexity is widely regarded as a major and enduring challenge of software engineering~\cite{sommerville2012large, pressman2014software, bass2021software, brooks1987no, benaroch2022much}.
Among the paradigms proposed to address it, Model Driven Engineering (MDE) has shown particular promise for improving development efficiency and consistency~\cite{schmidt2006model, brambilla2017model, hutchinson2011model, di2022correction}, and a wide range of MDE tools have been adopted across industries~\cite{steinberg2008emf, kolovos2008epsilon, miyazawa2019robochart, parracalderon2024ide4icds, cooper2021model, cederbladh2024early, viyovic2014sirius, wei2024towards}.
Such tools let engineers model different lifecycle stages (e.g.\ requirements and architecture), levels of abstraction, and viewpoints (e.g.\ safety and security).

However, the resulting models are typically built with different tools and are therefore \textit{heterogeneous}~\cite{bruneliere2020scalable}, which makes them hard to integrate and interoperate -- especially when requirements change -- and the impact on other models must be assessed: a manual, time-consuming, and error-prone process~\cite{briand2003impact, aizenbud2006model}.
The lack of automated change management is twofold: there is no holistic approach for complete model-based systems engineering in which models are produced at all key stages~\cite{bucchiarone2020grand}, and it is difficult to establish \textit{executable}, fine-grained (model-element-level) traceability links among heterogeneous models.

Existing MDE work largely \textit{maintains} or \textit{recovers} traceability links within semi-homogeneous environments---e.g.\ reactive links~\cite{ractiu2024using, ratiu2022reactive}, traceability recovery~\cite{delucia2008adams}, trace-link evolution~\cite{rahimi2016evolving}, cross-tool traceability~\cite{ratiu2023taming}, and Digital-Twin consistency management~\cite{muctadir2024consistency}.
A significant gap remains in the \textit{automated generation} of fine-grained, executable traceability links across heterogeneous models in a complete model-based process.

These challenges are particularly acute for safety-critical hardware-software systems such as Cyber-Physical Systems (CPSs) and Robotics and Autonomous Systems (RASs), where standards such as IEC 61508~\cite{iec61508} require software safety analysis and verification.
Arguing such safety often demands formal notations to specify software behaviour and verify properties such as deadlock freedom and convergence---yet formal methods have steep learning curves and are costly to maintain~\cite{woodcock2009formal}, and requirement changes force formal models to be reassessed, further compounding the loss of efficiency.

Recently, Large Language Models (LLMs) have attracted significant interest across software engineering~\cite{terragni2025future}, including requirements engineering~\cite{ronanki2024requirements}, code generation~\cite{gu2024effectiveness}, software testing~\cite{wang2024software}, program repair~\cite{fan2023automated}, and domain-specific modelling~\cite{camara2023assessment}.
Whilst promising results have been reported, a fundamental question remains: \textit{Can LLMs be leveraged, in conjunction with MDE, for a complete model-based (software) system engineering approach, in which the majority of models (including executable traceability models) can be automatically generated from a carefully engineered requirement model?}

To answer the above question, in this paper we propose RADIANT (Requirement-driven, Agent-Directed generatIon of trAceable, heterogeNeous sysTem models), an engineering methodology that combines MDE and Multi-Agent LLMs to support complete model-based (software) system engineering.
The core idea is that, given a carefully engineered requirement model, RADIANT (and its supporting tool) can automatically generate accurate and robust heterogeneous system models\footnote{By ``models'', we mean the \textit{metamodel}s for different aspects of the system and the \textit{model}s that conform to such \textit{metamodel}s.} together with executable traceability models containing model-element level \textit{executable} traceability links, by leveraging MDE and Multi-Agent LLM technologies.

We detail the phases of RADIANT and its supporting tool, and illustrate them on a case study engineering an Autonomous Underwater Vehicle (AUV): from a carefully engineered requirement model, RADIANT generates (i) a system concept model, (ii) a Domain Specific Modelling Language (DSML), (iii) a conforming system model, (iv) a formally verified behaviour model (automatically checked for deadlock freedom and convergence), and (v) element-level traceability models linking requirements to all generated artefacts.
Our evaluation shows reliable gains in artefact validity (with model-dependent semantic accuracy), an order-of-magnitude ($10$--$15\times$) efficiency gain over manual modelling, and transfer of the unmodified pipeline to a second domain.

\textbf{Contributions.} This paper makes the following contributions:
\begin{itemize}
    \item \textbf{A requirement-model-driven methodology for complete model-based engineering (central contribution).}
RADIANT combines MDE with Multi-Agent LLMs to drive a \emph{complete} model-based engineering process from a single carefully engineered requirement model: the system concept model, the domain-specific modelling language (DSML), the conforming system model, and the behaviour model are all generated automatically, rather than authored by hand.
    \item \textbf{A generic Multi-Agent LLM layer, and an analysis of when its decomposition helps (enabling contribution).}
Each RADIANT phase is realised by an instance of a generic Multi-Agent LLM layer of cooperating generate/check/refactor/trace agents and a check-gated repair agent; the layer is defined once, varies only in its per-phase agents and inputs, and is reused unchanged across domains.
Against a single-agent baseline, the decomposition reliably improves the \emph{validity} of the generated formal artefacts---and their \emph{executability} where the model's code generation permits---with additional, model-dependent gains in semantic accuracy.
    \item \textbf{Executable traceability by construction with a delivered change-impact facility (enabling contribution).}
Every layer emits a traceability model linking requirements to the model elements they produce across heterogeneous formalisms.
On top of these models, RADIANT provides an automated change-impact analysis facility that, given a requirement change, computes the affected model elements across all generated artefacts---a delivered capability, not merely the soft links of prior work.
    \item \textbf{Generation of formally verifiable behaviour models with an automated verify--repair loop (safety-critical contribution).}
RADIANT generates safety-critical behaviour models (RoboChart state machines~\cite{miyazawa2019robochart}) that plug directly into the established RoboChart verification toolchain~\cite{foster2020formal}: each generated state machine is automatically translated into CSP (Communicating Sequential Processes) and checked with the FDR refinement checker (e.g.\ for deadlock freedom and convergence), and models that fail are repaired using the returned counterexamples.
Our contribution is closing the loop from a requirement model to a formally verifiable behaviour model; on the case study, the generated model is automatically shown deadlock-free and convergent.
    \item \textbf{A comparative empirical evaluation and an open replication package.}
We evaluate RADIANT's accuracy, traceability, efficiency (a six-participant comparative study), and cross-domain applicability across three frontier LLMs (one open-weight, two proprietary); all artefacts, agent configurations, prompts, and evaluation data are publicly available.
\end{itemize}

The rest of the paper is organised as follows.
Section~\ref{sec:background} presents the three pillars our approach rests on and the running case study; Section~\ref{sec:approach} presents RADIANT and its supporting tool MALCOM; Section~\ref{sec:eval} is a feasibility demonstration on the case study, answering five research questions; Section~\ref{sec:discussion} discusses implications, generalisability, threats to validity, and related work; Section~\ref{sec:agenda} sets out a research agenda; and Section~\ref{sec:conclusion} concludes.

\section{Background}
\label{sec:background}

This section introduces the three pillars on which RADIANT rests---Model-Driven Engineering, Multi-Agent Large Language Models, and executable traceability with automated change management---and then presents the case study used throughout the rest of the paper.
We provide conceptual background here; technical details of the supporting tool are in Section~\ref{sec:implementation}.

\subsection{Model-Driven Engineering}

Model-Driven Engineering (MDE) is a paradigm in which models are first-class artefacts that \textit{drive} the engineering process~\cite{brambilla2017model, lara2017posteriori, schmidt2006model}: they are used to analyse, simulate and reason about a system and to generate part of its implementation, improving efficiency, consistency and quality assurance~\cite{volter2013model, hutchinson2011model}.
Two activities are central to MDE.
(1) \textit{Domain-Specific Modelling} lets experts model a system at an appropriate level of abstraction using Domain-Specific Modelling Languages (DSMLs), each realised as a \textit{metamodel}---for example, an Ecore metamodel built with the Eclipse Modelling Framework (EMF)~\cite{steinberg2008emf}---that defines its abstract syntax~\cite{kleppe2008software}.
(2) \textit{Model Management} operations---validation~\cite{kolovos2009on, warmer2003object}, model-to-model~\cite{kolovos2008epsilon,jouault2006atl} and model-to-text~\cite{rose2008epsilon} transformation---are then applied iteratively to generate artefacts (code, documentation) until the system is delivered.

An MDE process is only as good as the requirements that drive it.
Requirements Engineering (RE) underpins project success, and in safety-critical industries the \textit{Requirement Specification} can be contractually binding; yet RE remains poorly tooled---61\% of companies rely solely on word processors and spreadsheets~\cite{martins2018requirements}, making requirement-to-system traceability hard to establish.
A model-based approach to requirements is therefore a promising foundation for automation and traceability.

\subsection{Multi-Agent Large Language Models}

Large Language Models (LLMs) are transformer-based neural networks trained on large text corpora~\cite{chang2024survey}; modern LLMs~\cite{ouyang2022training, achiam2023gpt} perform \emph{in-context learning}---adapting to a task from instructions and examples in the prompt rather than parameter updates---and are typically aligned with human intent via Reinforcement Learning from Human Feedback (RLHF)~\cite{ouyang2022training, stiennon2020learning}.
Rather than fine-tuning a model for each task, LLMs can be steered at inference time through \textbf{prompt engineering}~\cite{brown2020language}.
Two strategies are central to RADIANT's layers: \textbf{few-shot prompting}~\cite{brown2020language}, which adds a handful of input--output examples to guide output format and reasoning, and \textbf{Chain-of-Thought (CoT) prompting}~\cite{wei2022chain}, which elicits intermediate reasoning steps for complex, multi-step tasks.

A single agent is nonetheless limited, as realistic tasks span multiple concerns.
Attention has thus shifted to \textbf{Multi-Agent LLM (MALLM)} frameworks~\cite{wu2023autogen, hong2023metagpt, guo2024multiagent}, which orchestrate several specialised agents that interact in a defined order and share a dialogue history.
Several target software engineering---CAMEL~\cite{li2023camel} (complementary roles), MetaGPT~\cite{hong2023metagpt} (an \textit{assembly line} of roles), and AutoGen~\cite{wu2023autogen} (a unified conversation framework integrating LLMs, tools, and human oversight); we build on AutoGen.

\subsection{Traceability and Automated Change Management}

Requirements rarely stay fixed, and every change must be propagated to the artefacts it affects.
\emph{Change impact analysis}---assessing a change's risk and cost before it is made---is therefore essential~\cite{lawrence1998software, sommerville2011software} and is mandated by functional-safety standards such as IEC 61508~\cite{iec61508} and ISO 26262~\cite{iso26262}, yet remains labour-intensive: nearly 65\% of practitioners regard managing requirement changes as a significant challenge~\cite{umar2023automated}.

Disciplined change management rests on \emph{traceability}---\textit{the ability to interrelate any uniquely identifiable software engineering artefact to any other, maintain required links over time, and use the resulting network to answer questions of both the software product and its development process}~\cite{cleland2012software}.
In practice, requirements and their links are managed in dedicated requirements-management tools such as IBM DOORS~\cite{doors} and Jama Connect~\cite{jamaconnect}, which maintain requirements and traceability within a single environment but do not automatically produce the cross-model, element-level links needed to follow a requirement change into heterogeneous downstream models.
MDE is better suited to such traceability, since model elements and their relationships are explicit, and a substantial body of work \emph{maintains} or \emph{recovers} links across engineering models---for example, reactive links that propagate changes~\cite{ractiu2024using, ratiu2022reactive} and traceability recovery~\cite{delucia2008adams, rahimi2016evolving} (Section~\ref{sec:related}).
These approaches, however, typically assume a common or semi-homogeneous environment, or rely on manual link specification; when artefacts are genuinely heterogeneous, tools often fall back on ``soft links''~\cite{denney2018tool} that point only to a referenced model rather than to the affected element, leaving the impact analysis to be done by hand.

These limitations motivate \emph{executable} traceability.
We call a traceability link \emph{executable} when it carries enough information for model-management operations (in particular, model-to-model transformations and model queries) to load and locate its source and target elements automatically---even across models defined in different DSMLs and technologies---so that change-impact analysis can be performed without manual intervention.

Such traceability is especially valuable for safety-critical systems, whose safety must be justified, often in an assurance case~\cite{wei2019model, kelly1999arguing} relating hazardous behaviours to the safety requirements that mitigate them~\cite{foster2020formal}.
Higher assurance levels typically demand formalising both the requirements and the software's behaviour---to prove properties such as deadlock freedom and convergence~\cite{foster2019isabelle}---which is itself hard to keep traceable unless the requirements are formalised too.

\subsection{Case Study}
\label{sec:case-intro}
We use an externally authored case study throughout this paper: an Autonomous Underwater Vehicle (AUV) of the National Oceanography Centre, previously modelled in RoboChart and verified in Isabelle/SACM~\cite{foster2020formal, wei2024access}, to which we apply RADIANT end-to-end in Section~\ref{sec:eval}.
The AUV is a portable, untethered remotely operated vehicle with a visual mapping system and verified on-board autonomy, used for light intervention tasks such as cathodic-protection surveys and coring.
As it operates under regulatory oversight, assuring its software-controlled autonomous behaviour is vital.

The AUV is modelled using RoboChart~\cite{foster2020formal} as a robotic platform (\textit{AUV\_Platform})---an abstraction layer over the hardware providing shared sensor, actuator and event variables---composed with an operator (\textit{AUV\_Operator}) that issues task instructions, an autopilot (\textit{AUV\_Autopilot}) that drives the actuators, and a Last Response Engine (LRE, \textit{LRE\_Ctrl}) that sits between them to enforce safety.
We detail the LRE and its behaviour, the focus of our Phase~5 demonstration, in Section~\ref{sec:demo}.




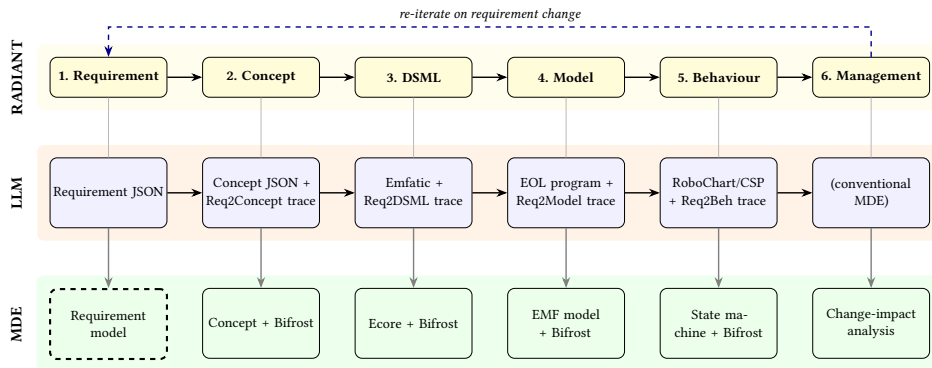
\begin{figure}[h!]
    \centering
    \resizebox{0.82\textwidth}{!}{%
    \begin{tikzpicture}[
        font=\small,
        phase/.style={draw, rounded corners, fill=yellow!18, minimum height=8mm,
                      text width=22mm, align=center, inner sep=2pt, font=\small\bfseries},
        art/.style={draw, rounded corners, minimum height=14mm, text width=22mm,
                    align=center, inner sep=2pt, fill=blue!6},
        mdl/.style={draw, rounded corners, minimum height=14mm, text width=22mm,
                    align=center, inner sep=2pt, fill=green!8},
        flow/.style={-{Stealth[length=2.2mm]}, thick},
        drop/.style={-{Stealth[length=2.2mm]}, thick, gray},
        conn/.style={gray!55, thin},
        iter/.style={-{Stealth[length=2.2mm]}, thick, dashed, blue!55!black},
    ]
    \node[phase] (p1) {1. Requirement};
    \node[phase, right=7mm of p1] (p2) {2. Concept};
    \node[phase, right=7mm of p2] (p3) {3. DSML};
    \node[phase, right=7mm of p3] (p4) {4. Model};
    \node[phase, right=7mm of p4] (p5) {5. Behaviour};
    \node[phase, right=7mm of p5] (p6) {6. Management};

    \node[art, below=12mm of p1] (a1) {Requirement JSON};
    \node[art, below=12mm of p2] (a2) {Concept JSON + Req2Concept trace};
    \node[art, below=12mm of p3] (a3) {Emfatic + Req2DSML trace};
    \node[art, below=12mm of p4] (a4) {EOL program + Req2Model trace};
    \node[art, below=12mm of p5] (a5) {RoboChart/CSP + Req2Beh trace};
    \node[art, below=12mm of p6] (a6) {(conventional MDE)};

    \node[mdl, dashed, very thick, below=12mm of a1] (m1) {Requirement model};
    \node[mdl, below=12mm of a2] (m2) {Concept + Bifrost};
    \node[mdl, below=12mm of a3] (m3) {Ecore + Bifrost};
    \node[mdl, below=12mm of a4] (m4) {EMF model + Bifrost};
    \node[mdl, below=12mm of a5] (m5) {State machine + Bifrost};
    \node[mdl, below=12mm of a6] (m6) {Change-impact analysis};

    \foreach \i/\j in {1/2,2/3,3/4,4/5,5/6} { \draw[flow] (p\i) -- (p\j); }
    \foreach \i in {1,2,3,4,5,6} { \draw[conn] (p\i) -- (a\i); }
    \foreach \i/\j in {1/2,2/3,3/4,4/5,5/6} { \draw[flow] (a\i) -- (a\j); }
    \foreach \i in {1,2,3,4,5,6} { \draw[drop] (a\i) -- (m\i); }

    \begin{scope}[on background layer]
        \node[fill=yellow!8,  rounded corners, fit=(p1)(p6), inner sep=2.5mm] (bandr) {};
        \node[fill=orange!10, rounded corners, fit=(a1)(a6), inner sep=2.5mm] (bandp) {};
        \node[fill=green!7,   rounded corners, fit=(m1)(m6), inner sep=2.5mm] (bandj) {};
    \end{scope}

    \node[font=\bfseries, rotate=90] at ([xshift=-4mm]bandr.west) {RADIANT};
    \node[font=\bfseries, rotate=90] at ([xshift=-4mm]bandp.west) {LLM};
    \node[font=\bfseries, rotate=90] at ([xshift=-4mm]bandj.west) {MDE};

    \draw[iter] (p6.north) -- ++(0,6mm) -| (p1.north);
    \node[font=\small\itshape] at ($(p1.north)!0.5!(p6.north)+(0,8.5mm)$) {re-iterate on requirement change};
    \end{tikzpicture}%
    }
\caption{Overview of RADIANT and its realisation in our tool MALCOM (Section~\ref{sec:implementation}). The middle (LLM) lane
produces JSON/source artefacts per phase via multi-agent generation; the bottom (MDE) lane transforms each into models and \textit{Bifrost} traceability.
Only the requirement model (Phase~1, dashed) is authored manually; all other artefacts are generated.}
    \label{fig:approach}
\end{figure}

\section{Proposed Approach}
\label{sec:approach}
This section presents RADIANT, our methodology, which combines MDE with Multi-Agent LLMs into a model-based engineering process driven by a single requirement model.
Figure~\ref{fig:approach} gives an overview in three lanes, the top one listing the six RADIANT phases.
The middle (LLM) lane shows the artefacts that the Multi-Agent LLM layers produce for Phases~2--5: JSON and source code, each paired with a requirement-to-artefact (\textit{Req2X}) trace.
The bottom (MDE) lane shows the domain models and the traceability models derived from these artefacts, ending in Phase~6 with model-management operations.
Each traceability model connects the requirements to the downstream model elements through executable, element-level links that drive automated change-impact analysis when requirements evolve.

\subsection{The Multi-Agent LLM Layer}
\label{sec:agentlayer}

\begin{figure}[b!]
    \centering
    \resizebox{0.68\linewidth}{!}{%
    \begin{tikzpicture}[
        font=\small,
        box/.style={draw, rounded corners=3pt, align=center, inner sep=3pt, line width=0.4pt},
        proxy/.style={box, text width=40mm, minimum height=9mm, fill=orange!12, draw=orange!55},
        mgr/.style={box, text width=48mm, minimum height=9mm, fill=green!12, draw=green!50},
        agent/.style={box, text width=26mm, minimum height=13mm, fill=blue!8, draw=blue!45},
        pipe/.style={-{Stealth[length=2.4mm]}, thick, black!65},
        sched/.style={-{Stealth[length=2.4mm]}, dashed, black!55},
        fb/.style={-{Stealth[length=2.4mm]}, dashed, red!65!black},
    ]
    \node[proxy] (proxy) {\textbf{User Proxy}\\[1pt]{\scriptsize provides requirements (JSON)}};
    \node[mgr, below=6mm of proxy] (mgr) {\textbf{Group Chat Manager}\\[1pt]{\scriptsize schedules agents round-robin}};

    \node[agent, below=16mm of mgr, xshift=-46.5mm] (gen) {\textbf{Generator}\\[1pt]{\scriptsize Extraction / Creation}};
    \node[agent, right=5mm of gen] (chk) {\textbf{Checker}\\[1pt]{\scriptsize corrects errors}};
    \node[agent, right=5mm of chk] (ref) {\textbf{Refactorer}\\[1pt]{\scriptsize (optional)}};
    \node[agent, right=5mm of ref] (trc) {\textbf{Trace Gen.}\\[1pt]{\scriptsize emits \textit{Req2X} links}};

    \begin{scope}[on background layer]
        \node[draw=black!30, dashed, rounded corners=5pt, fill=black!3,
              fit=(gen)(trc), inner sep=5mm] (group) {};
    \end{scope}
    \node[anchor=north west, font=\scriptsize\itshape, text=black!55]
          at ([xshift=2mm,yshift=-2mm]group.north west) {Group chat};

    \draw[pipe] (proxy) -- (mgr);
    \draw[sched] (mgr) -- (group.north);
    \draw[pipe] (gen) -- (chk);
    \draw[pipe] (chk) -- (ref);
    \draw[pipe] (ref) -- (trc);
    \draw[fb] (chk.south) -- ++(0,-3mm) -| (gen.south);
    \node[font=\scriptsize, text=red!65!black] at ($(gen.south)!0.5!(chk.south)+(0,-4.5mm)$) {correct};

    \node[box, fill=yellow!14, draw=yellow!55!black, text width=42mm, minimum height=10mm,
          below=11mm of group.south] (verify) {\textbf{Check gate} (MALCOMj)\\[1pt]{\scriptsize structural, traceability, FDR checks}};
    \node[agent, right=12mm of verify] (repair) {\textbf{Repair Agent}\\[1pt]{\scriptsize fixes the artefact}};
    \draw[pipe] (group.south) -- node[right=1pt, font=\scriptsize] {output} (verify.north);
    \draw[fb] ([yshift=2mm]verify.east) -- node[above, font=\scriptsize, text=red!65!black] {fail} ([yshift=2mm]repair.west);
    \draw[fb] ([yshift=-2mm]repair.west) -- node[below, font=\scriptsize] {repaired} ([yshift=-2mm]verify.east);
    \node[font=\scriptsize, below=6mm of verify.south, anchor=north] (out) {\textit{pass}: artefact $+$ \textit{Req2X} trace};
    \draw[pipe] (verify.south) -- (out.north);
    \end{tikzpicture}%
    }
    \caption{The generic Multi-Agent LLM layer, instantiated once per generative phase (Phases~2--5), as described in the text.}
    \label{fig:agentlayer}
\end{figure}
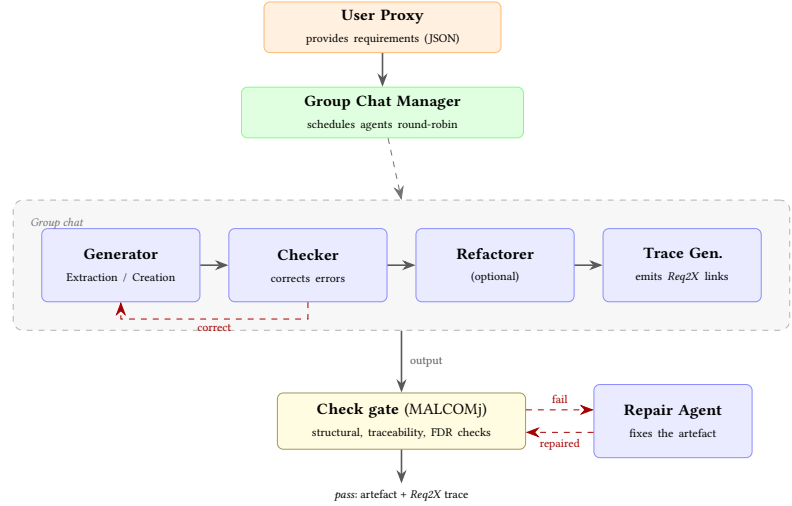

Phases~2--5 each generate a downstream artefact through a layer of collaborating LLM agents.
The layer is generic, but each generative phase instantiates its own agents, configured with phase-specific prompts and Chain-of-Thought exemplars; the agent configurations and the inputs each agent receives likewise change from phase to phase (detailed in Section~\ref{sec:engphases}).
Figure~\ref{fig:agentlayer} shows its structure.

A \textit{User Proxy} supplies the requirements (in JSON) to the layer as a single batch.
A \textit{Group Chat Manager} then schedules the agents round-robin, with all agents sharing a single dialogue history.
A \textit{Generator}---guided by few-shot and Chain-of-Thought prompting (Section~\ref{sec:background})---drafts the artefact; a \textit{Checker} validates it and, on error, hands control back to the Generator; an optional \textit{Refactorer} consolidates the result; and a \textit{Trace Generation Agent} emits the requirement-to-artefact (\textit{Req2X}) links.

Beyond the group chat, each layer's output passes a check gate (Section~\ref{sec:implementation}).
On a hard failure, a dedicated \textit{Repair Agent}---one per phase, configured for that phase's artefact---corrects the output and the check repeats, up to a retry cap.
The layer thus yields a checked artefact together with its traceability links.

\subsection{The Engineering Phases}
\label{sec:engphases}
We now describe the six phases, focusing on what is specific to each.
Phases~1 and~6 frame the generative core: Phase~1 produces the requirement model that drives the rest, and Phase~6 returns to conventional MDE for model management and change-impact analysis.
The core is Phases~2--5, each instantiating the generic layer of Figure~\ref{fig:agentlayer}. The \textit{Generator} is realised as an \emph{Extraction} agent for the concept and DSML phases (2--3) and a \emph{Creation} agent for the model and behaviour phases (4--5); the \textit{Refactorer} is added in Phases~3 and~4, and every phase runs the \textit{Checker}, \textit{Trace Generation} and \textit{Repair} agents.
Each generator receives the requirement model and the outputs of earlier phases (the \textit{Concept} model, the DSML, and the system model), plus, for Phase~5, the target behaviour grammar.
Each of these phases also defines its own \textit{Repair Agent}, which the check gate (Section~\ref{sec:agentlayer}) invokes on a hard failure to correct that phase's artefact---respectively the Concept model, the Emfatic metamodel, the EOL program, and the RoboChart state machine.

\subsubsection{Phase 1: Requirement Modelling}
\textbf{Phase 1} is the only manual step: the engineer authors the requirement model.
In this work we use a lightweight requirement DSML\footnote{The metamodel is available at: \url{https://github.com/wrwei/RADIANT/tree/main/malcom.requirement/malcom.requirement.model/metamodel}.} that captures \textit{Requirements} (user, functional and non-functional) together with a \textit{Glossary}, \textit{Stakeholder}s and \textit{Categor}ies, organised into \textit{RequirementPackage}s for modularity.

From our experience, a few conventions in writing requirements lead to markedly more accurate generation.
First, developers distinguish \textit{Concept}s from their \textit{Instance}s---in MDE terms, \textit{meta-elements} and \textit{instance objects}---so that the Multi-Agent LLM layers can exploit this structure.
Concretely, \textit{Concept}s are capitalised and \textit{Instance}s enclosed in quotation marks within each requirement statement.
Second, requirements are kept atomic, with complex requirements broken into individual ones, which helps the LLMs interpret them.
Finally, and optionally, requirements are modularised, giving the LLMs a sense of system boundaries and encouraging a modular design.

With the requirement model in place, a model-to-text (M2T) transformation serialises it into JavaScript Object Notation (JSON), a format that is easy to exchange and for LLMs to process (Section~\ref{sec:implementation}).
The approach is not tied to our DSML: any requirement model with a corresponding M2T transformation to this JSON format can drive the pipeline.


\subsubsection{Phase 2: Concept Extraction}
\textbf{Phase 2} extracts the \textit{Concept}s and \textit{Instance}s implicit in the requirements, so that later phases can reuse this structure.
The \textit{Extraction Agent} identifies the \textit{Concept}s and \textit{Instance}s, the \textit{Checker Agent} confirms them, and the \textit{Trace Generation Agent} emits the requirement-to-concept links.

The layer's two JSON outputs---the \textit{Concept}s/\textit{Instance}s and the traceability links---are transformed into models conforming to two DSMLs we provide: a \textit{Concept} DSML, and \textit{Bifrost}, a traceability metamodel we created\footnote{The \textit{Bifrost} metamodel is available at: \url{https://github.com/wrwei/RADIANT/tree/main/malcom.bifrost/malcom.bifrost.model/metamodel}.} (Section~\ref{sec:implementation}).
Each \textit{Bifrost} \textit{Trace} links a \textit{source} and a \textit{target} through \textit{ExternalReference}s carrying the \textit{version} (for change detection) and \textit{location} of the traced element, realising the \emph{executable} traceability defined in Section~\ref{sec:background}.

\subsubsection{Phase 3: DSML Generation}

\textbf{Phase 3} generates a DSML from the requirement model and the \textit{Concept} model.
This phase applies when no suitable domain language yet exists; when one is already available, Phase~3 can be bypassed and the existing DSML supplied directly to Phase~4.

Since LLMs are strongest at generating text, the DSML is produced as source code that yields the metamodel when processed.
Several textual notations exist (e.g.\ Emfatic~\footnote{\url{https://eclipse.dev/emfatic/}.} for EMF Ecore, or the SysML~v2 textual notation~\footnote{\url{https://github.com/Systems-Modeling/SysML-v2-Pilot-Implementation}}); we generate Emfatic, which is parsed into an Ecore metamodel (Section~\ref{sec:implementation}).
The \textit{Refactoring Agent} restructures the generated code---for instance, eliciting inheritance and hoisting shared attributes to parent classes.
The phase yields the Emfatic source, the Ecore metamodel, the \textit{Req2DSML} trace, and the \textit{Bifrost} traceability model (Table~\ref{tab:phases}).

\subsubsection{Phase 4: Model Creation}

\textbf{Phase 4} creates a system model that conforms to the DSML generated in Phase~3.

The generators receive the requirements, the \textit{Concept} model (Phase~2), and the DSML (Phase~3).
Rather than emitting the model directly in an exchange format such as XMI~\footnote{\url{https://www.omg.org/spec/XMI/}}, the agents generate, check, and refactor a program in the Epsilon Object Language (EOL)~\cite{kolovos2006epsilon} that is executed to construct the model programmatically (Section~\ref{sec:implementation})---source code being both abundant in LLM training data and directly consumable by MDE tools.
The phase yields the EOL program, the resulting EMF model, the \textit{Req2Model} trace, and the \textit{Bifrost} traceability model (Table~\ref{tab:phases}).

\subsubsection{Phase 5: Behaviour Modelling}
\textbf{Phase 5} is required for safety-critical systems and optional for non-safety-critical engineering: it generates lower-level safety-related models---for example, safety-analysis models (e.g.\ Failure Mode and Effects Analysis~\cite{wei2022designing}), software behaviour models as state machines~\cite{foster2020formal}, or assurance cases~\cite{wei2019model}.
We focus on behaviour models as state machines, which formal-method tools can check for properties such as deadlock freedom and convergence.

The \textit{Behaviour Model Creation Agent} receives the requirements, the \textit{Concept} model (Phase~2), the DSML (Phase~3), the system model (Phase~4), and the target behaviour grammar.
The phase yields the behaviour-model source (a RoboChart/CSP state machine, automatically verified as described in Section~\ref{sec:implementation}), the \textit{Req2Beh} trace, and the \textit{Bifrost} traceability model (Table~\ref{tab:phases}).

\subsubsection{Phase 6: Model Management}
\textbf{Phase 6} returns the developer to conventional MDE.
With these models in hand, the developer continues with the model-management operations of a typical MDE process---validation, model-to-text and model-to-model transformation, and so on---until the system is delivered.

Crucially, when the requirements change, an automated change-impact analysis can be performed that identifies which downstream elements are affected.
At any stage, developers can re-iterate: the DSML is generated once, a minor requirement change may only re-run Phases~4--5, and a fundamental change re-runs from Phase~2 or~3.
One current limitation is that RADIANT does not yet propagate \emph{manual} edits of the generated models back to the requirements or other artefacts (round-trip engineering); in the interim, the \textit{Bifrost} models flag which artefacts a requirement change affects, and full round-trip support is left to future work.

\subsection{Implementation}
\label{sec:implementation}
We now detail MALCOM (\emph{Multi-Agent LLM CO-modelling and co-management environment}), the model-based tool that supports RADIANT.

Three principles shape MALCOM.
(1) \textit{Separate domain assets from generic layers}: all case-study-specific content (requirements, system descriptions, prompt assets) lives in a per-case-study directory, while the Multi-Agent LLM layers are entirely generic.
(2) \textit{Generate, check, refactor, then trace}: every layer follows this same pipeline, repairing its artefact on a failed check before accepting it (Section~\ref{sec:agentlayer}).
(3) \textit{Traceability by construction}: every layer produces a \textit{Bifrost} model so that change-impact analysis is available throughout, not bolted on afterwards.

MALCOM comprises two parts mirroring the LLM and MDE lanes of Figure~\ref{fig:approach}.
\textit{MALCOMp} is the Python implementation, which builds the Multi-Agent LLM layers (Figure~\ref{fig:agentlayer}) on the AutoGen framework.
\textit{MALCOMj} is the Java/Eclipse implementation, which runs the model-to-model and model-to-text transformations and constructs the Ecore, EMF, and RoboChart models; a Sirius-based environment additionally supports manual requirement modelling in Phase~1.
An optional browser front-end, MALCOM-web\footnote{Available at: \url{https://github.com/wrwei/RADIANT/tree/main/MALCOM-web}.}, exposes the same pipeline interactively through a FastAPI/WebSocket server that streams the multi-agent dialogue; it introduces no new methodology and was not used for the reported results.
MALCOMj builds on a set of established MDE technologies:
\begin{itemize}
\item \textit{Eclipse Modelling Framework (EMF)}~\cite{steinberg2008emf} provides \textit{Ecore}, the metamodelling language in which we define DSMLs.
\item \textit{Eclipse Epsilon}~\cite{kolovos2006eclipse} manages models across arbitrary modelling technologies; its \textbf{Emfatic} syntax expresses Ecore metamodels as text.
\item \textit{Eclipse Sirius}~\cite{viyovic2014sirius} builds graphical modelling editors over EMF through \textit{ViewPoints}.
\item \textit{RoboTool}~\cite{miyazawa2019robochart} supports graphical modelling, validation, and the generation of mathematical definitions for model checking; its \textit{RoboChart} notation covers both architecture and state machines.
\end{itemize}

Table~\ref{tab:phases} maps each RADIANT phase to its MALCOMp layer, MALCOMj transformation, and output artefacts.

\begin{table}[tbp]
\centering
\caption{RADIANT phases and their realisation in MALCOM. Each Multi-Agent LLM layer (MALCOMp) produces JSON/source artefacts that MALCOMj transforms into models and \textit{Bifrost} traceability.}
\label{tab:phases}
\resizebox{\textwidth}{!}{%
\begin{tabular}{llll}
\hline
\textbf{Phase} & \textbf{MALCOMp layer (agents)} & \textbf{MALCOMj transformation} & \textbf{Output artefacts} \\ \hline
2 & Concept Extraction (Extraction, Checker, Trace, Repair) & JSON$\to$Concept; JSON$\to$Bifrost & Concept/Instance JSON, Concept model, Req2Concept, Bifrost model \\
3 & DSML Generation (Extraction, Checker, Refactorer, Trace, Repair) & Emfatic$\to$Ecore; JSON$\to$Bifrost & Emfatic source, Ecore metamodel, Req2DSML, Bifrost model \\
4 & Model Creation (Creation, Checker, Refactorer, Trace, Repair) & EOL execution; JSON$\to$Bifrost & EOL program, EMF model, Req2Model, Bifrost model \\
5 & Behaviour Modelling (Creation, Checker, Trace, Repair) & RoboChart build/FDR; JSON$\to$Bifrost & RoboChart/CSP source, state machine, Req2Beh, Bifrost model \\ \hline
\end{tabular}%
}
\end{table}

In Phase~5, the agents generate RoboChart behaviour models---a CSP-based notation~\cite{brookes1984theory, foster2020formal} implemented in Xtext~\cite{bettini2016implementing}---from the RoboTool grammar.
These are automatically translated to CSP and discharged with the FDR refinement checker~\footnote{FDR: The CSP Refinement Checker, \url{https://cocotec.io/fdr/index.html}.} for properties such as deadlock freedom and convergence, with failing models repaired from the returned counterexamples---a translate-and-repair loop we adapt from~\cite{forge}.
For the change-impact analysis of Phase~6, a content hash of each requirement is compared against the \textit{version} attribute recorded in the corresponding \textit{Bifrost} \textit{ExternalReference}; a mismatch flags the requirement as changed, and the affected downstream elements are read off the traceability links.
More sophisticated change detection (e.g.\ via Git history or model indexing~\cite{barmpis2013hawk}) is left to future work.

\section{Feasibility Demonstration}
\label{sec:eval}

We evaluate RADIANT on the AUV case study introduced in Section~\ref{sec:case-intro} as a feasibility demonstration: the contribution is the requirement-driven, multi-agent generation methodology, and the aim is to show that it can produce accurate, traceable models end-to-end, to isolate the effect of the multi-agent decomposition, and to characterise where that effect is---and is not---realised across different LLMs.
We pose five research questions.

\subsection{Research Questions}
\label{sec:eval:rqs}
\begin{description}
    \item[\textbf{RQ1: Accuracy.}] How accurate are RADIANT's generated artefacts (concept, DSML, model, behaviour) against an expert reference?
    \item[\textbf{RQ2: Multi-agent necessity.}] Does the generate--check--refactor--trace multi-agent decomposition improve accuracy over a single-agent baseline?
    \item[\textbf{RQ3: Traceability.}] Are the generated executable trace links correct and complete, and do they support automated change-impact analysis?
    \item[\textbf{RQ4: Efficiency.}] How much development effort does RADIANT save versus manual modelling with the same MDE tooling?
    \item[\textbf{RQ5: Generality.}] Does the same unchanged infrastructure apply across requirement models from different domains?
\end{description}

\subsection{End-to-End Demonstration on the AUV}
\label{sec:demo}
For this case study, we create an Eclipse modelling project that is publicly available~\footnote{All AUV artefacts referenced below (requirement model, the generated JSON, Concept/DSML/model/behaviour outputs, traceability transformations, and the EMF/RoboChart models) are in the public repository at \url{https://github.com/wrwei/RADIANT/tree/main/case_studies/auv}.}. We refer to the artefacts under this project in subsequent phases.


In \textbf{Phase~1} we author the requirement model in \textit{MALCOMj} and export it to JSON via a model-to-text transformation. \textbf{Phase~2} feeds this JSON to \textit{MALCOMp}'s Concept extraction, yielding a Concept JSON and a Req2Concept trace, which \textit{MALCOMj} transforms into Concept and \textit{Bifrost} models.
\textbf{Phase~3} provides the requirement and Concept JSON to the DSML layer, producing an Emfatic metamodel and a Req2DSML trace, assembled into an Ecore metamodel and a \textit{Bifrost} model.
\textbf{Phase~4} provides the requirement JSON and Emfatic source to the model-creation layer, producing an EOL program and a Req2Model trace; executing the EOL yields an EMF model conforming to the metamodel, together with its \textit{Bifrost} model.
To fully exploit RoboChart and demonstrate interoperability, we then add a model-to-text transformation from the EMF model to a RoboChart model.

For RADIANT \textbf{Phase 5}, since the AUV is safety critical, we demonstrate how state machines can be generated and formally verified by FDR.
In this work, we focus on the behaviour model of the ``Last Response Engine'' (LRE), which provides run-time safety assurance for the AUV.
Architecturally, the LRE sits between the operator and autopilot components.
The operator, which can be a human or the navigation system, provides instructions to the LRE to support execution of tasks, such as requesting a particular heading and velocity.
The autopilot controls the AUV actuators, and takes advice only from the LRE.
The LRE’s job is to avoid hazardous behaviours, such as getting too close to an obstacle, or entering ``Object Proximity Exclusion Zones'' (OPEZ), and engaging evasive manoeuvres if necessary.
The LRE functions in four modes: in Operator Control Mode (OCM) it is inactive, passing control inputs through to the autopilot; in Main Operating Mode (MOM) it takes control for normal operation at full speed; in High Caution Mode (HCM), nearing an obstacle, it drops velocity; and in Collision Avoidance Mode (CAM), on a potential collision, it manoeuvres away.
The LRE has several high-level safety requirements, a number of examples from~\cite{foster2020formal} are:
\begin{itemize}
    \item The LRE shall enter MOM from OCM when the following conditions hold: (i) the velocity is less than 0.1ms$^{-1}$; (ii) the distance to a static obstacle is > 300mm; (iii) the distance to a dynamic obstacle is greater than 7500mm; (iv) the operator requests it; and (v) the AUV is not in an OPEZ.
    \item On entering MOM, the LRE shall advise a velocity of 1ms$^{-1}$.
\end{itemize}

\begin{figure}[tbp]
    \centering
    \includegraphics[width=.8\linewidth]{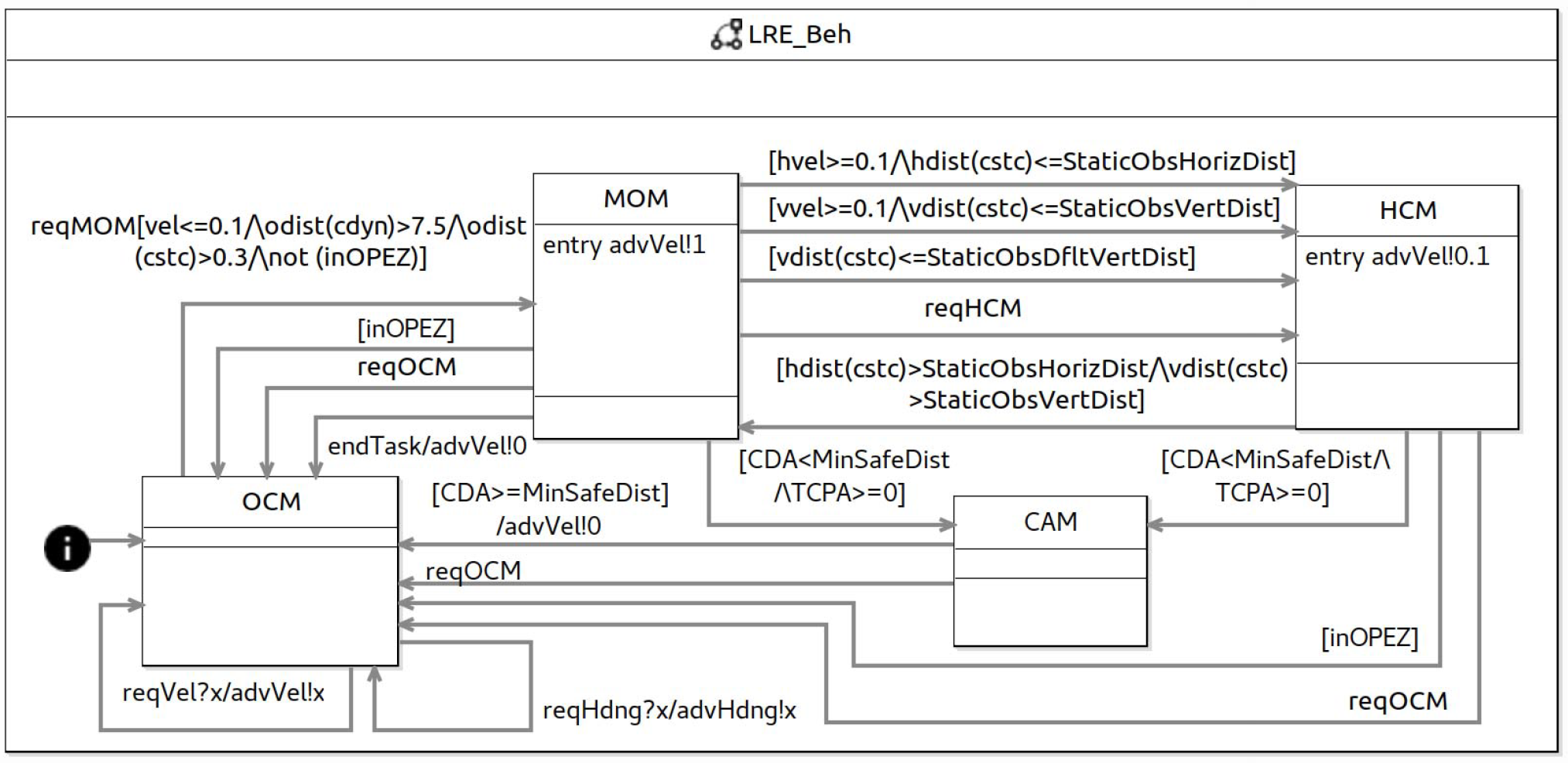}
    \caption{The state machine of the Last Response Engine modelled using RoboChart~\cite{foster2020formal}.}
    \label{fig:LRE_STM}
\end{figure}

The states of LRE and the safety requirements can be expressed using a state machine, as shown in Figure~\ref{fig:LRE_STM}.
Our aim in this phase is to generate the state machine from the requirement JSON in Phase 1.

The state machine uses variables such as \textit{cobs} (closest obstacle), \textit{inOPEZ} (whether the AUV is in an Object Proximity Exclusion Zone), \textit{hvel}/\textit{vvel} (horizontal/vertical velocity), \textit{cstc}/\textit{cdyn} (closest static/dynamic obstacle) and \textit{depth}; constants such as \textit{MinSafeDist}; and functions \textit{odist}/\textit{hdist}/\textit{vdist}, which return the (horizontal/vertical) distance from the AUV to a given \textit{Obstacle}.

In the requirement model we represent each variable, constant and function as a single requirement element, and each state and each inter-state transition condition as a single element (unlike the higher-level safety requirements listed above~\cite{foster2020formal}).
With \textit{MALCOMp} we generate a CSP state-machine descriptor matching Figure~\ref{fig:LRE_STM}.

In \textbf{Phase~6} we perform conventional MDE activities on the generated models: FDR checks the behaviour state machine for deadlock freedom and convergence~\cite{foster2020formal}; an Epsilon Validation Language (EVL) validation can be run whenever the model changes; and automated change-impact analysis, on an injected requirement change, identifies the affected elements across the generated models.
As these are standard model-management activities, we omit the details.

\subsection{Evaluation Settings}\label{sec:eval:settings}
\subsubsection{LLMs setting.}In MALCOMp the LLM is a swappable component, not the contribution.
To test whether the pipeline generalises beyond a single model, we run it unchanged across three frontier models from different families and providers: one open-weight model, DeepSeek-V4~\cite{liu2024deepseek_v3}, and two proprietary, API-only models, QWEN3-max (Alibaba)~\cite{yang2025qwen3technicalreport} and Claude Sonnet~4.6 (Anthropic)~\cite{anthropic2025claudesonnet}.
DeepSeek-V4 and QWEN3-max are run at temperature~0.0; for Claude the temperature is omitted, as the Anthropic API deprecates it for this model.
To account for the stochasticity of LLM generation, each (model, configuration) cell is repeated ten times.

\subsubsection{Statistical tests.}For each LLM we run ten Single-agent and ten Multi-agent repetitions, and compare the two configurations stage by stage, where a \emph{stage} is one of the five accuracy metrics reported in Table~\ref{tab:accuracy}.
The two sets are independent samples (not matched pairs), and the per-stage scores are not normally distributed (Shapiro--Wilk); we therefore use the Mann--Whitney U (rank-sum) test for the continuous F1 stages and Fisher's exact test for the two binary-rate stages, all two-sided.

\subsubsection{Metrics.}Since RADIANT generates multiple artefacts, we score four representative ones against the expert reference: the Concept model (JSON), the DSML metamodel (Emfatic), the model-creation code (EOL) together with the EMF model it builds, and the behaviour model (RoboChart, checked via its CSP semantics).

For the \emph{Concept model}, we compute a set-based precision/recall/F1 over the extracted vocabulary facts $\{$Concept, Instance, Instance\_of$\}$ against the reference, using alias-aware canonicalisation (lower-casing and non-alphanumeric stripping) so auditable semantic equivalences are matched --- a transparent count of which concepts and instances were correctly extracted, missed, or hallucinated.

For the \emph{DSML} (the Emfatic metamodel), we follow the matching strategy of prior work~\cite{chen2023comparative} and compute precision, recall and F1 over its key elements (classes, attributes and references): a true positive is a correctly matched element, a false positive an unmatched generated element, and a false negative a missed reference element.
Each element is scored $S(x)=1$ for an exact or semantically-equivalent match, $0.5$ for a partial match, and $0$ otherwise.
The matching rules are: (1) two \emph{classes} score $1$ if their names are identical or semantically equivalent, and $0.5$ if partially matched (one a subclass of the other, or with substantially different members); (2) two \emph{attributes} match if their parent classes match and their names are identical or semantically similar (e.g.\ \texttt{onSickLeave} and \texttt{isSick}); (3) two \emph{references} match whenever the association is correctly identified, regardless of whether it is declared \texttt{ref} (non-containment) or \texttt{val} (containment), since requirements rarely distinguish the two.

For the \emph{model-creation code} (EOL) we do not score code similarity; instead we \emph{execute} it.
Each run's generated Emfatic is converted to Ecore and the EOL program is run against it with the Epsilon engine; we then check that the resulting instance \emph{conforms} to the metamodel using EMF's generic \texttt{Diagnostician}, and score the produced model against the reference with set-based precision/recall/F1 over objects, attributes and links --- measuring whether the generated code actually builds a conformant model, not its textual resemblance.

For the \emph{behaviour model} we compute set-based structural F1 over the state machine's states, initial node and transition edges (multiplicity-aware, so a dropped parallel transition is penalised), together with per-transition trigger, guard and action facts, against the reference; we additionally check deadlock-freedom and convergence with FDR.
We report the structural metrics because the pipeline deliberately abstracts concrete guards into boolean variables for FDR-verifiability, so a full-fidelity semantic comparison would measure that intended abstraction rather than generation quality.

\begin{table}[tbp]
\caption{Accuracy of generated artefacts against the hand-authored expert reference, Single- vs Multi-agent, for three LLMs (mean over ten repetitions on the AUV case study). Concept and DSML are set-based F1; \emph{DSML syntax} is the fraction of syntactically-valid Emfatic; \emph{Model/EOL} is the fraction of runs whose EOL executes to a metamodel-conformant model; \emph{Behaviour} is multiplicity-aware structural F1 over the state machine.}
\label{tab:accuracy}
\centering
\small
\begin{tabular}{llccccc}
\hline
\textbf{Model} & \textbf{Config} & \begin{tabular}[c]{@{}c@{}}Concept\\ vocab F1\end{tabular} & \begin{tabular}[c]{@{}c@{}}DSML\\ syntax-valid\end{tabular} & \begin{tabular}[c]{@{}c@{}}DSML\\ class F1\end{tabular} & \begin{tabular}[c]{@{}c@{}}Model/EOL\\ exec.\ conform\end{tabular} & \begin{tabular}[c]{@{}c@{}}Behaviour\\ struct.\ F1\end{tabular} \\ \hline
DeepSeek-V4 & Single & 0.90 & 0.50 & 0.93 & 0/10 & 0.97 \\
DeepSeek-V4 & Multi & \textbf{0.95} & \textbf{1.00} & 0.91 & \textbf{10/10} & \textbf{1.00} \\ \hline
QWEN3-max & Single & 0.84 & 0.00 & 0.88 & 0/10 & \textbf{1.00} \\
QWEN3-max & Multi & 0.80 & \textbf{0.80} & \textbf{0.90} & \textbf{4/10} & 0.98 \\ \hline
Claude Sonnet 4.6 & Single & 0.89 & 1.00 & 0.92 & 0/10 & 0.97 \\
Claude Sonnet 4.6 & Multi & 0.88 & 1.00 & 0.86 & 0/10 & 0.97 \\ \hline
\end{tabular}
\end{table}
\subsection{RQ1: Accuracy}
For accuracy, we conduct pair-wise comparisons on the AUV case study against a hand-authored expert reference (the AUV Concept model, DSML, EMF model-creation EOL, EMF model, and behaviour model), produced by reasoning over the requirements and the MDE tooling without invoking any LLM\footnote{The requirement model is created by us and stays the same.}.
For each of the three LLMs we repeat RADIANT ten times in two configurations: a \textit{Single agent} baseline (one direct call per layer using that layer's primary agent) and the default \textit{Multi agent} pipeline (each layer running the generate--check--refactor--trace agents of Section~\ref{sec:agentlayer}).
Each generated artefact is scored against the reference using the metrics defined above.

The results are summarised in Table~\ref{tab:accuracy}.
Across the three models a consistent split emerges: the multi-agent decomposition reliably improves the \emph{syntactic and structural validity} of the downstream formal artefacts, whereas its effect on the \emph{semantic accuracy} of the upstream extraction is model-dependent.

\paragraph{DeepSeek-V4 --- the multi-agent decomposition helps at every stage.} Concept vocabulary F1 rises from $0.90$ (single) to $0.95$ (multi); Emfatic syntactic validity rises from $0.50$ to $1.00$ as the check/refactor agents repair malformed metamodels; the generated EOL goes from never instantiating a conformant model ($0/10$ single) to building one in all ten runs ($10/10$ multi); and the behaviour state machine's structural F1 rises from $0.97$ to $1.00$ as the requirement-completeness gate recovers transitions the boolean abstraction would otherwise merge.
Every difference favours the multi-agent configuration.

\paragraph{QWEN3-max --- decisive gains on validity and executability, neutral on semantics.} The starkest validity effect appears here: every single-agent run emits an Emfatic metamodel that the Eclipse parser rejects (syntactic validity $0/10$), because the single agent omits the mandatory \texttt{package} declaration; the multi-agent check/refactor loop repairs the header and lifts validity to $0.80$.
Conformant-model construction likewise improves from $0/10$ (single) to $4/10$ (multi): the multi-agent loop closes several of the EOL's feature-reference errors, though the remaining failures --- where the generated EOL writes to RoboChart feature names its own metamodel never declares --- are a residual code-generation weakness the decomposition cannot fully eliminate.
On the semantic stages, by contrast, multi-agency is neutral: Concept vocabulary F1 is marginally lower under multi ($0.84 \to 0.80$, the collaborative loop slightly over-produces the vocabulary) and behaviour structural F1 is already near-perfect single-agent ($1.00$ vs $0.98$).

\paragraph{Claude Sonnet~4.6 --- semantic gains absent and code generation the limiting factor.} Sonnet's check/refactor loop tends to over-produce, so on the DSML stage the multi-agent class F1 sits slightly below the single agent ($0.92 \to 0.86$) rather than improving on it, and the behaviour structural F1 merely ties ($0.97$).
Most strikingly, Sonnet's EOL does not reliably execute --- $0/10$ conformant models for \emph{both} configurations --- failing with parse errors and feature references on the wrong class, a genuine code-generation weakness rather than a measurement artefact; here the decomposition has no executable artefact to improve.

In short, the multi-agent decomposition reliably improves artefact \emph{validity} across models---and their \emph{executability} where the model's code generation permits---while its effect on semantic accuracy is model-dependent --- positive for DeepSeek-V4, neutral for QWEN3-max, absent for Sonnet~4.6, whose weak EOL generation also caps the pipeline regardless of configuration.
We test these differences statistically in RQ2.

\subsection{RQ2: Multi-Agent Necessity}
This question isolates whether the multi-agent decomposition is \emph{necessary}, comparing the Multi- against the Single-Agent control over the ten repetitions per stage with the tests of Section~\ref{sec:eval:settings} (Mann--Whitney U for continuous F1, Fisher's exact for the binary-rate stages).

\begin{figure}[tbp]
    \centering
	\includegraphics[width=\linewidth]{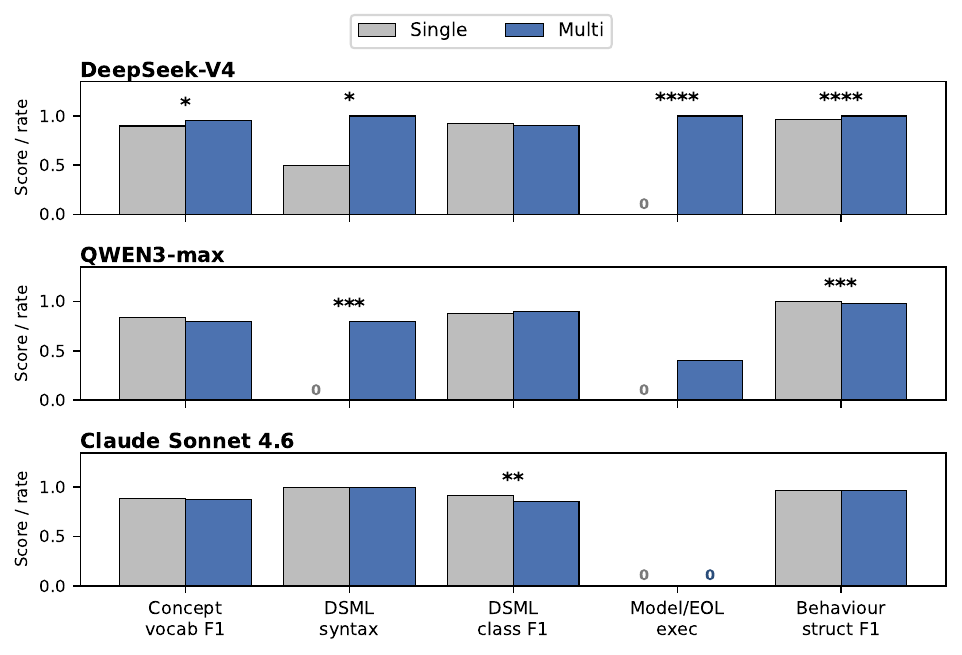}
\caption{Per-stage Single- vs.\ Multi-agent scores per model (mean over ten repetitions). Stars mark significant differences in \emph{either} direction (Mann--Whitney U for continuous F1; Fisher's exact for binary rates): $p<0.05$: *, $p<0.01$: **, $p<0.001$: ***, $p<0.0001$: ****. Bar heights show direction; a ``0'' is a measured zero (e.g.\ Sonnet's $0/10$ EOL runs), not missing data.}
	\label{fig:analysis}
\end{figure}

The answer separates cleanly by the \emph{kind} of artefact (Figure~\ref{fig:analysis}), and that is the key result.
The one effect that is \textbf{significant and in the multi-agent's favour for every model whose single agent leaves it imperfect} is \emph{DSML syntactic validity}: $0.50 \to 1.00$ for DeepSeek-V4 (Fisher $p=0.033$) and $0.00 \to 0.80$ for QWEN3-max ($p=7.1\times10^{-4}$); Sonnet~4.6 is already $1.00$ single-agent, so there is nothing to recover.
Conformant-model \emph{construction} improves wherever the model's code generation makes it attainable, but its significance tracks the model: decisive for DeepSeek-V4 ($0/10 \to 10/10$, $p=1.1\times10^{-5}$), a positive but \emph{non-significant} trend for QWEN3-max ($0/10 \to 4/10$, $p=0.087$ at ten repetitions), and unattainable for Sonnet~4.6 ($0/10$ under both).
On the \textbf{semantic} stages the picture is \emph{model-dependent}: multi-agency significantly improves DeepSeek-V4 (Concept vocabulary $0.90 \to 0.95$, $p=0.038$; behaviour structural F1 $0.97 \to 1.00$, $p=3.3\times10^{-5}$), but for QWEN3-max and Sonnet~4.6 it does not improve Concept (both differences non-significant), and their only significant semantic movements run \emph{against} multi-agency --- QWEN3-max's behaviour F1 dips $1.00 \to 0.98$ ($p=7.1\times10^{-4}$, a small but consistent drop from an already-perfect single-agent baseline) and Sonnet~4.6's DSML class F1 falls $0.92 \to 0.86$ ($p=1.1\times10^{-3}$) as its check/refactor loop over-produces.
The remaining differences (e.g.\ DSML class F1 for DeepSeek-V4 and QWEN3-max) are not significant.

We therefore answer RQ2 as: the multi-agent decomposition's most robust, model-independent contribution is to the \emph{syntactic validity} of the generated formal artefacts --- it reliably repairs invalid metamodels, which is exactly the work of the check/refactor loop.
Its contribution to downstream \emph{executability} (conformant EMF models) is in the same direction but model-gated: significant only where the model's intrinsic code generation is strong enough (DeepSeek-V4), a non-significant trend for QWEN3-max, and unattainable for Sonnet~4.6. Its contribution to upstream \emph{semantic} accuracy is model-dependent and can even be mildly negative when a model's check/refactor loop over-produces.
The benefit is thus a property of the model--pipeline interaction, not of the decomposition in isolation.

\subsection{RQ3: Traceability}
\label{sec:eval:trace}
Unlike approaches that recover trace links post hoc, every RADIANT phase emits a \textit{Bifrost} model linking each requirement to the elements it produces, so traceability is available \emph{by construction}.
We assess it along two axes: the \emph{correctness and completeness} of the generated links, and their ability to \emph{support automated change-impact analysis}.
Both are scored offline over the artefacts already produced for RQ1, with no further LLM calls.

Traceability is itself a capability of the decomposition: trace links are emitted by the dedicated trace agent in each Multi-Agent layer, whereas the Single-Agent baseline produces no resolvable trace (only an unchecked concept listing).
The numbers below thus characterise the default (Multi-Agent) pipeline, and the single-agent comparison is qualitative.

\paragraph{Link correctness and completeness.} We score three quantities over the ten Multi-Agent runs per model.
\emph{Accuracy}: for the behaviour layer, the generated (requirement, transition-edge) links reproduce the hand-authored expert trace (18 transitions) exactly (F1~$=1.00$ for all three models, matched on the edge since textual labels are immaterial); for the concept layer, the concept extraction's requirement-linked micro-F1 ($0.53$--$0.72$) \emph{is} the requirement$\rightarrow$concept link accuracy.
\emph{Resolvability} (the fraction of links whose target is actually present in the artefact) is $1.00$ for concept, DSML and model across all models, and for behaviour $1.00$/$0.97$/$0.73$ (DeepSeek/QWEN/Sonnet) --- the generation weakness that depresses Sonnet's executable artefacts also leaves more of its behaviour links unresolved.
\emph{Coverage} (requirements with $\geq 1$ link) is $\geq 0.98$ for the architecture-side layers; the one gap is behaviour ($0.67$ for DeepSeek and QWEN), as the behaviour trace links transitions but not the variable, constant and state requirements.

\paragraph{Change-impact correctness.} Exercising the delivered facility (Section~\ref{sec:implementation}) end-to-end, we inject each of the eighteen behaviour requirements as a change and compare the impact set computed from each run's trace graph against the expert-intended impact.
Over $180$ changes per model it attains precision, recall and F1 of $1.00$ --- flagging exactly the affected transition.
Injecting the root architecture requirement instead propagates across the concept, DSML and model layers (three phases), demonstrating the cross-phase digital thread.
The mechanism is exact; its practical reliability is bounded only by link resolvability (hence model-dependent), not by the computation.

\begin{table}[H]
\caption{Pre-study background survey responses from the six participants (P1--P6); each item was rated on a four-point experience scale: 1~=~\emph{novice} (0--6 months), 2~=~\emph{intermediate} (6 months--1 year), 3~=~\emph{experienced} (1--3 years), 4~=~\emph{expert} (${>}3$ years).}
\label{tab:survey}
\centering
\small
\begin{tabular}{clcccccc}
\hline
\textbf{\#} & \textbf{Experience with\ldots} & \textbf{P1} & \textbf{P2} & \textbf{P3} & \textbf{P4} & \textbf{P5} & \textbf{P6} \\ \hline
Q1 & large language models (e.g.\ ChatGPT, Claude) & 2 & 1 & 2 & 3 & 2 & 3 \\
Q2 & using LLMs to assist programming or modelling & 2 & 1 & 2 & 3 & 2 & 3 \\
Q3 & model-driven engineering (MDE / MBSE) & 3 & 3 & 4 & 2 & 4 & 2 \\
Q4 & Eclipse Emfatic or Ecore modelling & 3 & 3 & 4 & 2 & 4 & 2 \\
Q5 & the Eclipse platform & 3 & 3 & 4 & 2 & 4 & 2 \\
Q6 & formal methods (e.g.\ model checking, theorem proving) & 1 & 3 & 3 & 1 & 3 & 1 \\ \hline
\end{tabular}
\end{table}

\subsection{RQ4: Efficiency}
To evaluate efficiency, we invited six participants and surveyed their experience with LLMs, MDE, and formal methods (Table~\ref{tab:survey}). Given their diverse backgrounds, a warm-up phase of tutorials and a workshop gave all participants a comparable understanding. Each participant performed the task in both the manual and automated settings; to mitigate training effects we used a crossover design, one group manual-first and the other automated-first. The six span intermediate-to-expert MDE and novice-to-experienced formal-methods experience; Table~\ref{tab:efficiency} reports all of them, and we highlight two contrasting profiles---P1 (experienced in MDE but a novice in formal methods) and P2 (experienced in both).

\begin{table}[tbp]
\caption{Comparative experiment for efficiency evaluation. All participants start from the same given requirement model (the input to RADIANT); the automated setting fixes the pipeline to DeepSeek-V4.}
\centering
\begin{tabular}{lccccc}
\hline
\multirow{2}{*}{Participant} & \multicolumn{5}{c}{Time Spent (in minutes)/Accuracy (in percentage)} \\ \cline{2-6}
                             & \begin{tabular}[c]{@{}c@{}}AUV\\ DSML\end{tabular}
                             & \begin{tabular}[c]{@{}c@{}}EMF\\ Model\end{tabular}
                             & \begin{tabular}[c]{@{}c@{}}RoboChart\\ Arch.\end{tabular}
                             & \begin{tabular}[c]{@{}c@{}}Behaviour\\ Model\end{tabular}
                             & \begin{tabular}[c]{@{}c@{}}Change Impact\\ Analysis\end{tabular} \\ \hline
P1 (Man.)  & 50/88\%   & 160/86\%  & 60/92\%  & 240/91\% & 60/N/a       \\
P2 (Man.)  & 36/88\%   & 123/87\%  & 70/94\%      & 120/94\%    & 45/N/a   \\
P3 (Man.)  & 27/95\%   & 99/93\%  & 60/93\%      & 117/93\%    & 35/N/a   \\
P4 (Man.)  & 39/86\%   & 120/89\%  & 80/87\%      & 200/87\%    & 56/N/a   \\
P5 (Man.)  & 30/96\%   & 90/93\%  & 60/93\%      & 130/93\%    & 59/N/a   \\
P6 (Man.)  & 60/84\%   & 153/86\%  & 90/77\%      & 200/77\%    & 62/N/a   \\
P1 (Auto.) & 8/95\%    & 12/97\%   & $<$1/97\%     & 12/96\%   & 1/N/a    \\
P2 (Auto.) & 6/95\%   & 18/97\%   & $<$1/97\%     & 8/97\%   & 1/N/a    \\
P3 (Auto.) & 8/97\%    & 10/97\%   & $<$1/96\%     & 5/97\%   & 1/N/a    \\
P4 (Auto.) & 6/95\%   & 18/96\%   & $<$1/96\%     & 10/97\%   & 1/N/a    \\
P5 (Auto.) & 3/97\%    & 11/97\%   & $<$1/96\%     & 5/96\%   & 1/N/a    \\
P6 (Auto.) & 6/93\%   & 20/97\%   & $<$1/97\%     & 15/97\%   & 1/N/a    \\
\hline
\end{tabular}
\label{tab:efficiency}
\end{table}

All participants started from the same given requirement model---the input to RADIANT---and produced the downstream AUV models of Section~\ref{sec:demo}, scored against our reference answers; the \textit{Concept} model is an internal intermediate the pipeline consumes, so it is not among the artefacts they author by hand.
We also asked the participants to perform a change impact analysis on the models based on one change in requirement.

In the manual setting, the participants used only the provided MDE tooling (MALCOMj), together with the AUV description, the LRE state machine (Figure~\ref{fig:LRE_STM}) and a short RoboChart tutorial.
The per-task times and accuracies are in Table~\ref{tab:efficiency}: accuracy is each artefact's element-level F1 against the expert reference (the metric of RQ1) as a percentage. 
Two effects stand out beyond the numbers.
P1, less experienced with the formalisms, had to revisit the DSML and EMF model after discovering design flaws (reflected in their 50- and 160-minute totals) and spent 240 minutes against the RoboChart manual to build the behaviour state machine --- versus P2's 120. 
Both authored their RoboChart models by hand-writing Epsilon Generation Language (EGL) model-to-text transformations~\cite{rose2008epsilon}.
For change impact, the participants were asked to add an attribute \texttt{rel\_dist} to the \texttt{Obstacle} composite type and traced its effects manually.

In the automated setting, the participants ran RADIANT via MALCOM on that same requirement model with the pipeline fixed to DeepSeek-V4 (the strongest model above), checking and lightly editing the generated artefacts in far less time; the RoboChart transformation is simply executed, not hand-written.
Because model and input are fixed, the generated artefacts are common across participants, so the automated times reflect review effort and the accuracies are the pipeline's rather than the individual's.
For the behaviour model, MALCOM generates a state machine matching the control-flow structure of Figure~\ref{fig:LRE_STM}; as its guards are abstracted to boolean variables for FDR-verifiability, participants checked the structure and reinstated the concrete guards.
For change impact, they only check the automated results.

Setting aside the model-to-text transformation, the automated process cuts every participant's effort by roughly an order of magnitude---from $278$--$510$ manual minutes to $20$--$42$ automated minutes, a $10$--$15\times$ reduction (mean ${\approx}12.5\times$), at no cost in accuracy ($93$--$97\%$ automated vs.\ $77$--$96\%$ manual).

\subsection{RQ5: Generality}

For generality, we applied RADIANT to a second, independently-authored case study from a different robotic domain: \emph{SRanger}\footnote{Available at: \url{https://github.com/wrwei/RADIANT/tree/main/case_studies/sranger}.}, a small reactive ground robot that turns in place on detecting an obstacle, with an IR distance sensor, a differential-drive actuator, and a three-mode controller (Moving, Turning, Final).
Its requirement model (23 requirements) was written for a separate code-generation study, not for RADIANT, and we reused the Multi-Agent layers of the AUV study \emph{unchanged}, including their prompts.

Running the full pipeline (phases~2--5) end-to-end, every phase passed its check gate without any layer or prompt change: Concept extraction produced a 9-concept model covering all requirements; DSML creation a syntactically valid 13-class Emfatic metamodel; the EOL executed to a metamodel-conformant EMF model; and the behaviour layer produced a RoboChart state machine with exactly the three modes and seven transitions specified, verified deadlock-free and convergent by FDR.
This second domain also revealed one trace-resolution gap we then fixed generally: concept names appearing in the requirement text only in lower-case or multi-word form were not matched; the AUV results are unaffected.

We frame this as evidence of \textit{applicability across domains}, not measured accuracy: no expert reference was authored for SRanger, so we report no accuracy numbers for it.
We claim only that the unmodified layers transfer and drive every phase --- including formal verification --- to a passing artefact.

\section{Discussion}
\label{sec:discussion}

\subsection{Implications}
RADIANT takes a position on how generative AI should enter model-based engineering: not a single LLM producing a finished system in one shot, but a chain of cooperating Multi-Agent LLM layers driven by a carefully engineered requirement model and managed with established MDE infrastructure.
Our three-model evaluation refines this position: the decomposition delivers a reliable, model-independent gain in the \emph{syntactic validity} of the formal artefacts --- precisely the work of the check/refactor loop --- and lifts their \emph{executability} where the model's code generation permits, while its effect on upstream \emph{semantic} accuracy is model-dependent: positive for a requirement-faithful model (DeepSeek-V4), neutral for QWEN3-max, and absent for one whose loop over-produces (Claude Sonnet~4.6), whose weak code generation also caps end-to-end viability regardless of configuration.
It also cuts development effort by an order of magnitude (10--15$\times$) over manual modelling and generates executable, element-level traceability by construction --- not recovered post hoc --- on which change-impact analysis is exact.
We regard this as an honest characterisation of \emph{when} the architecture helps, rather than a universal claim.

\subsection{Design Rationale}
We made three deliberate design choices.
(1) Splitting each phase into generate/check/refactor/trace roles lets each agent focus on a narrow task, the gain RQ2 isolates.
(2) Emitting a JSON intermediate keeps the LLM in a text-rich format and makes the pipeline generic --- any requirement model with a model-to-text transformation to the JSON schema can drive it.
(3) Generating Emfatic and EOL \emph{source} rather than model files targets artefacts both abundant in LLM training data and directly consumed by MDE tools, which then parse or execute them into conformant models, keeping the trust boundary on established tools.

\subsection{Generalisability}
Our generality evidence (RQ5) shows that the unchanged Multi-Agent LLM layers, including their prompts, transfer to a second, independently-authored case study and drive every phase to a passing artefact.
It does not establish broad \textit{domain} generality, which would require evaluation across many diverse domains, each with an independent reference; we treat that as future work rather than a claim.

\subsection{Related Work}
\label{sec:related}

In the domain of Model-Driven Engineering (MDE), several efforts\cite{alfonso2024besser,cabot2023lowcode} have been dedicated to improving the MDE process and enhancing its efficiency.
Our study is mainly relevant to the lines of research in LLMs for Model-Based System Engineering.

\textit{Model changes management. } In the context of traceability management, ~\cite{ractiu2024using,ratiu2022reactive} propose a reactive links approach to effectively propagate changes across heterogeneous engineering models.
Their method explicitly captures dependencies between model elements, enabling automatic detection and synchronization of changes.
And some solutions focus on requirements traceability like \cite{delucia2008adams} introduces a traceability recovery tool integrated within the ADAMS artefact management system and Eclipse IDE.
\cite{rahimi2016evolving} presents an automated approach to evolving traceability links between requirements and source code as software systems change over time.
\cite{ratiu2023taming} presents a comprehensive investigation into managing cross-tool traceability in safety-critical engineering domains.
They introduce a novel approach featuring a dynamic integration of trace information directly into artefact representations, allowing seamless navigation and maintenance across diverse engineering tools.
And in the context of consistency management, \cite{muctadir2024consistency} proposes a graph-based consistency management framework tailored for Digital Twin (DT) models.
They identify unique challenges associated with DT models, including heterogeneity, frequent evolution, and data intensity, and address them using a unified representation and graph databases.

\textit{LLMs4MDE} The construction of domain models leveraging LLM is explored in \cite{busch2023chatgpt}, \cite{camara2023assessment}, \cite{bell2024introducing}, \cite{netz2024natural}, and \cite{alshareef2023generative}.
\cite{chen2023comparative} conducts a comparative study on the use of large language models (LLMs) specifically GPT-3.5 and GPT-4, for fully automated domain modelling from natural language descriptions.
They found that although LLMs show strong potential, they still struggle with recall, especially in capturing relationships between domain elements.
\cite{bertram2022neural} presented a few-shot approach using GPT-based models to translate informal requirements into structured domain-specific languages (DSLs).
\cite{arulmohan2023domain} explores the use of Large Language Models (LLMs), particularly GPT-3.5, to extract domain models from textual requirements in agile product backlogs.
\cite{silva2025tree} introduces an innovative application of the Tree-of-Thoughts (ToT) framework for enhancing domain modelling using Large Language Models (LLMs).
They proposed decomposing the domain modelling process into structured, incremental phases, systematically generating and evaluating intermediate ``thoughts'' to improve the quality of modelling outcomes.

\textit{Multi-Agent LLMs.} AI agents based on large language models exhibit significant advantages in planning, decision-making, and tool invocation, effectively decomposing complex tasks into manageable subtasks.
Recent studies have explored the application of multi-agent large language models to software engineering tasks such as code review and requirements management, demonstrating preliminary success.
\cite{wang2024unity} proposes CoRe, an innovative approach utilizing collaborative agents based on large language models (LLMs) for recommending code reviewers.
Its agents take distinct roles (Coordinator, Code Analyzer, Reviewer Evaluator, and others) to improve recommendation accuracy and interpretability.
\cite{LLMs4RE4Collaboration} proposes MARE, a multi-agent collaboration framework leveraging Large Language Models (LLMs) to automate the entire Requirements Engineering (RE) process, including elicitation, modelling, verification, and specification.
The framework assigns dedicated roles to multiple agents (Stakeholder, Collector, Modeler, Checker, Documenter), each equipped with specific tasks and interactions via a shared workspace.

\subsubsection{Traceability links}

The reactive-links approach~\cite{ractiu2024using} noted above creates a traceability-link metamodel and uses model transformations to translate engineering artefacts into models conforming to it; however, it relies on models being stored in a common environment, which impedes its applicability.
A larger body of work instead \emph{recovers} trace links post hoc over existing artefacts, progressing from information-retrieval techniques~\cite{antoniol2002recovering} to deep-learning and pre-trained language-model approaches~\cite{guo2017semantically,lin2021traceability} (see~\cite{cleland2012software} for an overview).

In contrast to the above, RADIANT does not recover or maintain links within a homogeneous environment, nor generate a single model in isolation: it generates the full chain of heterogeneous models \textit{and} their executable, element-level traceability from one requirement model, using generic Multi-Agent LLM layers.

\subsection{Threats to Validity}
\label{sec:threats}
\textit{Internal validity.} LLM generation is stochastic, so a single run would not characterise accuracy or robustness; we therefore repeat RADIANT ten times per (model, configuration) and report distributions with statistical tests.
A second internal threat is metric validity: rather than opaque embedding similarity, our metrics are transparent and inspectable --- set-based precision/recall/F1 over extracted facts (with auditable alias matching), syntactic-validity checks, and, for the model-creation code, actual execution and EMF metamodel-conformance --- so a score reflects which elements were correctly produced, missed, or hallucinated.
The MDE transformations themselves are deterministic, so variability is confined to the generation phase.
RQ2's fifteen stage--model $p$-values are uncorrected: the strongest effects survive a conservative Bonferroni adjustment, the marginal ones ($p\approx0.03$--$0.04$) being indicative only.

\textit{External validity.} Our measured evaluation centres on a single case-study family (the AUV) and its behaviour controller, so broader system classes may exercise the layers differently; RQ5 partly mitigates this by transferring the unmodified pipeline to a second, independently-authored domain (SRanger).
Model coverage spans three frontier LLMs (one open-weight, two proprietary) but is not exhaustive; in particular the model-dependent finding motivates a broader model sweep as future work.
The efficiency study involves only six participants, so its absolute figures are indicative rather than definitive.
We mitigate these by grounding both case studies (the AUV and SRanger) in externally authored material, by evaluating both open and closed models, and by hedging the generality claim above; the residual construct risk is that the AUV \emph{reference} is hand-authored by us.

\subsection{Reproducibility and Artefact Availability}
RADIANT and all artefacts are open source\footnote{\url{https://github.com/wrwei/RADIANT}}: the requirement models, prompt and chain-of-thought assets, the Multi-Agent layer implementation (MALCOMp), the MDE transformations and metamodels (MALCOMj), the interactive UI, the per-model evaluation configurations, and the multi-model sweep results.
Any artefact can be regenerated from the requirement model and the reported metrics re-computed with the included evaluation harness.

\section{A Research Agenda for the Community}
\label{sec:agenda}
RADIANT answers an existence question---can a complete set of heterogeneous, traceable system models be generated from a single requirement model by combining MDE and Multi-Agent LLMs?---and in doing so surfaces follow-on questions for the wider community.

\textit{Compliance and quality across model families.} Our layers were tuned and evaluated on a fixed set of LLMs; how prompt assets, chain-of-thought guidance, and the generate--check--refactor decomposition behave across other model families and the many smaller open-weight models is an open, measurable question.

\textit{Broader and more diverse case studies.} Establishing domain generality, as opposed to the technical-space independence our generic layers target, requires evaluation across structurally different domains and system classes; community-contributed requirement models would accelerate this.

\textit{Deeper assurance-case integration.} The \textit{Bifrost} traceability produced by construction is a natural basis for structured assurance cases (GSN, SACM); wiring the generated models and their links into a machine-checkable assurance argument would extend RADIANT toward a certification-oriented route.

\textit{Fuller automation and additional formalisms.} Reducing the human review handoff, supporting additional DSMLs and downstream formalisms, and scaling change-impact analysis (e.g.\ via git history or model-indexing frameworks) each open a distinct line of work, with one generic front-end and several formal targets per system class.

\section{Conclusion}
\label{sec:conclusion}
We presented RADIANT, a requirement-model-driven methodology that combines MDE with Multi-Agent LLMs, and its supporting tool MALCOM.
From a single carefully engineered requirement model, RADIANT generates a system concept model, a domain-specific modelling language, a conforming system model, and a safety-critical behaviour model, each accompanied by executable, element-level traceability, without further manual authoring.

Our evaluation across three frontier models showed that the Multi-Agent LLM layers deliver a reliable, model-independent improvement in the \emph{validity} of the generated formal artefacts---and, where the model's code generation permits, their \emph{executability}---while their effect on upstream \emph{semantic} accuracy is \emph{model-dependent}: substantial for a requirement-faithful model (DeepSeek-V4), neutral for QWEN3-max, and absent for a model whose collaborative loop over-produces (Claude Sonnet~4.6).
We also showed that the approach improves development efficiency by roughly an order of magnitude ($10$--$15\times$) over manual modelling with the same tooling, and that the same layers apply unchanged to a second domain.

We view the contribution as a structural argument about where generative AI belongs in model-based engineering: not as a one-shot oracle, but as the generative end of a requirement-driven, multi-agent chain whose outputs are managed---and made traceable---by the MDE infrastructure the community already trusts.
This paves the way toward end-to-end, automated, and trustworthy model-based systems engineering.

\bibliographystyle{ACM-Reference-Format}
\bibliography{references}
\end{document}